\documentclass[11pt]{article}
\usepackage{moriond,epsfig}
\usepackage{graphicx}

\newcommand{\as}{\alpha_s}
\newcommand{\CA}{C_A}
\newcommand{\TR}{T_R}

\newcommand{\ie}{\emph{i.e.}\ }
\newcommand{\eg}{\emph{e.g.}\ }
\newcommand{\order}[1]{\mathcal{O}\left(#1\right)}

\begin{document}
\vspace*{4cm}
\title{INFRARED SAFE DEFINITION OF JET FLAVOUR}

\author{ G.P. SALAM }

\address{LPTHE, CNRS UMR 7589, \\
  Universit\'e P. et M. Curie (Paris VI) and Universit\'e
  Denis Diderot (Paris VII)\\
  75252 Paris cedex 05, France}

\maketitle\abstracts{ Though it is widely taken for granted that it
  makes sense to separately discuss quark and gluon jets, normal jet
  algorithms lead to a net parton-level jet flavour that is infrared
  (IR) unsafe. This writeup illustrates the problem and explains how
  the $k_t$ algorithm can be modified to provide an IR safe
  parton-level flavour. As well as being of use in theoretical
  calculations that require a projection of higher-order contributions
  onto a flavour-channel of the lowest order, jet-flavour algorithms
  also open up the prospect of significant improvements in the
  accuracy of heavy-quark jet predictions. }

\section{Introduction}

Over 350 articles on SPIRES refer in their title to ``quark-jet(s)''
or ``gluon-jet(s)''. This presupposes that such a distinction can be
made sensible at parton level. 

It is well known that there is no unique way of defining jets --- for
example the mapping of an $n+1$ parton configuration onto $n$-jets
(and the decision of whether that mapping should be carried out in the
first place) becomes ambiguous when all $n+1$ particles are hard and
widely separated in angle. This ambiguity persists when trying to
identify $n$ \emph{flavoured} jets from $n+1$ partons.  Nevertheless
one might hope that, if one identifies the flavour of a jet as the sum
of flavours of its constituent partons, then that flavour will be
meaningful, i.e.\ infrared (IR) safe, just as the energies and angles
of the jets are IR safe.

\begin{figure}[tbp]
  \centering
  \includegraphics[width=0.3\textwidth]{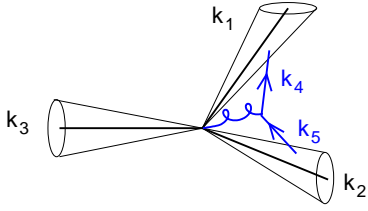}
  \caption{$5$ parton configuration clustered to 3 jets, where  a
    large-angle soft $q\bar q$ pair ($k_4, k_5$) contaminates the
    flavour of two of the jets.}
  \label{fig:softqqbar}
\end{figure}

When mapping $n+1$ partons onto $n$ jets, the IR safety of the flavour
holds trivially in normal jet algorithms. However with $n+2$ or more
partons, there can be configurations (fig.~\ref{fig:softqqbar}) with
an extra large-angle soft $q\bar q$ pair stemming from the branching
of a soft gluon, such that the quark is clustered into one jet while
the anti-quark is clustered into another. Both those jets have their
flavours `contaminated' by the soft (anti)quark. Because there is a
soft divergence associated with the production of the gluon that
branched to the large-angle $q\bar q$ pair, a perturbative calculation
of the jet flavours will lead to a result that diverges in the
infrared. This happens notably for cone, \cite{cone} $k_t$~\cite{kt}
and Cambridge~\cite{cam} type jet algorithms.

The jet algorithm that offers the most scope for resolving this
problem is the $k_t$ algorithm. It introduces a distance measure
$y_{ij}$ between each pair of particles $i$ and $j$, recombines the
pair of particles that are closest, and then repeats the procedure
until one has a predetermined number of jets (or all jets are
sufficiently widely separated). The $k_t$ distance measure (in its
form for $e^+e^-$ collisions), is
\begin{equation}
  \label{eq:yij-kt}
  y_{ij}^{k_t} = \frac{2\min(E_i^2, E_j^2)}{Q^2} (1 - \cos \theta_{ij})\,,
\end{equation}
where $E_i$ is the energy of particle $i$, $\theta_{ij}$ is the angle
between particles $i$ and $j$ and $Q$ is the centre of mass energy.
This choice of distance measure can be justified on the grounds that
the emission of a gluon is associated with two divergences: one as the
gluon energy vanishes, and another as the gluon becomes close in angle
to any other particle, for example:
\begin{equation}
  \label{eq:soft-col-mat}
  [dk_j] |M^2_{g\to g_i g_j}(k_j)| \simeq \frac{\as\CA}{\pi} \frac{dE_j}{\min(E_i,E_j)}
  \frac{d\theta_{ij}^2}{\theta_{ij}^2}\,, \qquad (E_j \ll E_i\,, \;\,
  \theta_{ij} \ll 1)\,.
\end{equation}
Such a structure involving two divergences holds only for gluon
production. Quark production in contrast has only a collinear
divergence:
\begin{equation}
  \label{eq:soft-col-mat-g2qq}
  [dk_j] |M^2_{g\to { q_i}\bar { q_j}}({ k_j})| \simeq
  \frac{\as\TR}{2\pi} 
  \frac{ dE_j}{{\max} (E_i,E_j)} 
  \frac{d\theta_{ij}^2}{\theta_{ij}^2}\,,\qquad\quad ({ E_j} \ll
  { E_i}\,, \;\,
  \theta_{ij} \ll 1)\,,
\end{equation}
(note the ``$\max$'' in the denominator).  When one is interested
mainly in the \emph{kinematics} of jets this is largely irrelevant, since
most of the branchings in an event produce gluons. However when
investigating flavour, a problem arises because the $y_{ij}^{kt}$
makes it easy for a soft quark to recombine with a hard particle even
though there is no corresponding divergence for the production of that
soft quark.

A simple solution to the problem is to modify the jet distance measure
to reflect the structure of divergences in the theory. So we introduce
a new ``flavour distance'',
\begin{equation}
  \label{eq:yij-flavour}
  y_{ij}^{F} = \frac{2(1-\cos\theta_{ij})}{Q^2} \times\left\{
    \begin{array}[c]{ll}
      \max(E_i^2, E_j^2)\,, & \quad\mbox{softer of $i,j$ is flavoured,}\\
      \min(E_i^2, E_j^2)\,, & \quad\mbox{softer of $i,j$ is flavourless.}
    \end{array}
  \right.
\end{equation}
\begin{figure}[htbp]
  \centering
  \includegraphics[width=0.95\textwidth]{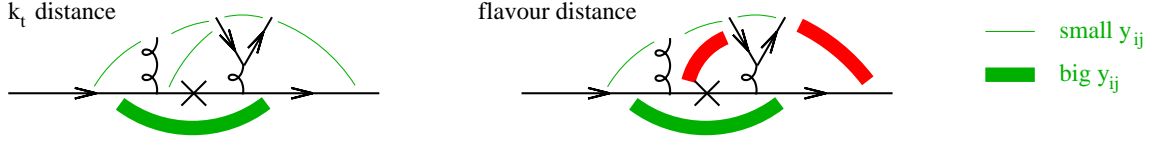}
  \caption{Representation of $y_{ij}$ distances between various
    particles in an $e^+e^-$ event with a $k_t$ distance measure and
    the flavour distance measure. Thick lines indicate a large
    $y_{ij}$, thin lines a small $y_{ij}$.}
  \label{fig:schematic-dists}
\end{figure}%
Note that this requires information on the flavour of each object, and
so can only be used at parton level (an exception is for heavy flavour
to which we return later). Figure~\ref{fig:schematic-dists}
illustrates the $y_{ij}$ distances between various partons in an event
with the two distance measures and in particular shows how a soft
quark has a large $y_{ij}^{F}$ with all hard particles in the event.
Thus it will first recombine with the antiquark of similar softness.
This will produce a gluon-flavoured object which can recombine with
hard objects without changing their flavour. So a modification of
the distance measure to better reflect the divergences in the theory
allows one to obtain an infrared-safe definition of jet
flavour.\cite{BSZFlavour}

\begin{figure}[htbp]
  \centering
  \includegraphics[width=0.47\textwidth]{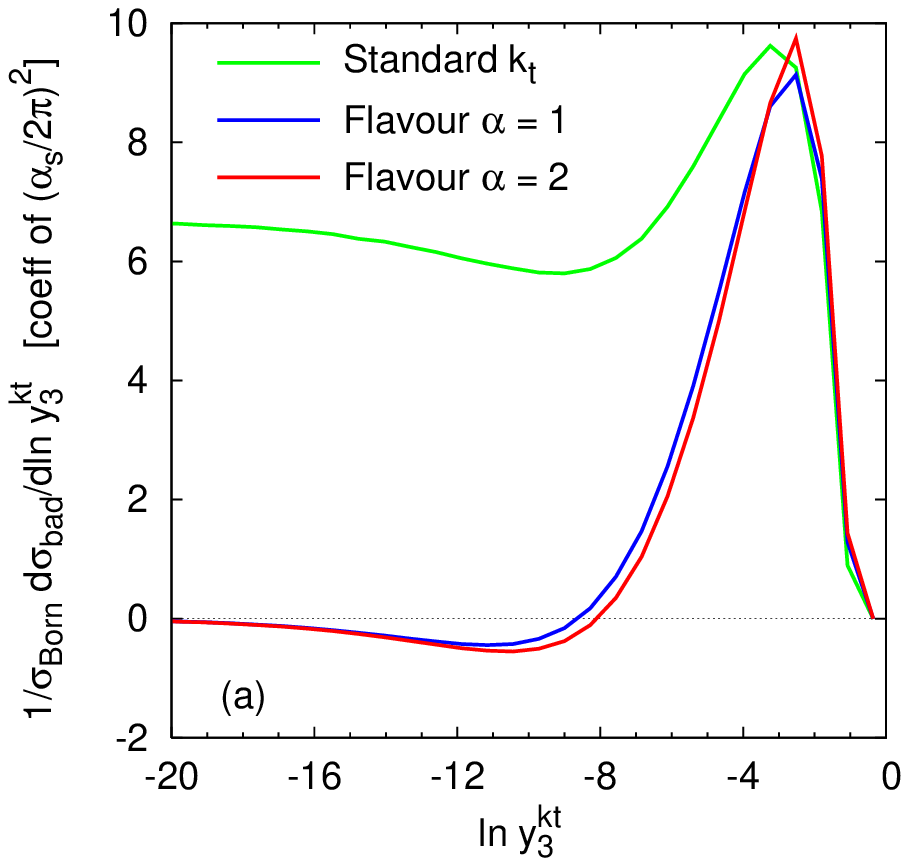}
  \hfill
  \includegraphics[width=0.477\textwidth]{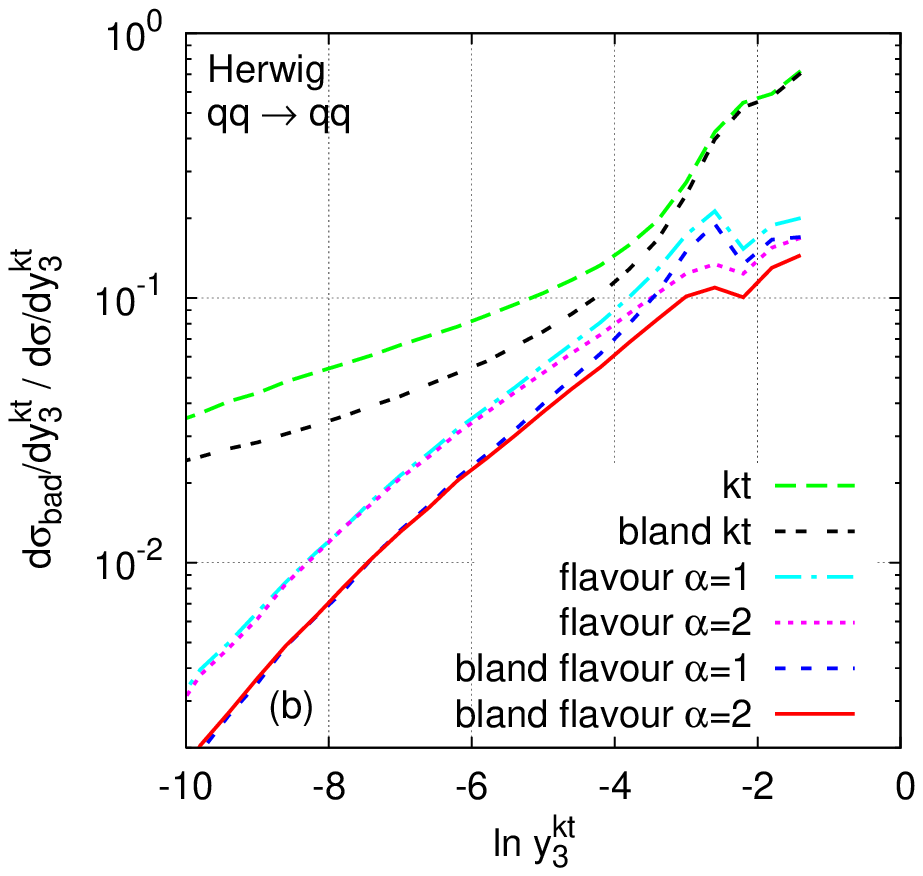}
  \caption{Proportions of events whose flavour is misidentified after
    clustering with various algorithms, as a function of the softness
    of the event: (a) in $e^+e^-$, coefficient of $\order{\as^2}$,
    calculated with EVENT2,\protect\cite{EVENT2} (b) in $qq \to qq$
    LHC events (with $\sim 1$~TeV jets) calculated using
    Herwig.\protect\cite{Herwig}}
  \label{fig:y3dists}
\end{figure}

The IR safety can be concretely illustrated by taking a fixed-order
parton-level $e^+e^-$ event, clustering it to two jets and examining the
cross section for cases where the flavour of the two jets is not
simply that of $q$, $\bar q$. If this cross section is plotted as a
function of $y_{3}^{kt}$, the $y_{cut}$ resolution threshold in the
$k_t$ algorithm above which the event is clustered to two jets, the
cross section should vanish as $y_{3}^{kt}\to 0$, \ie as the
soft/collinear limit is approached. At order $\as^2$,
fig.~\ref{fig:y3dists}a, this is seen to happen for the flavour
algorithms (which actually form a class defined by $\max(E_i^2,E_j^2)
\to [\max(E_i,E_j)]^{\alpha} \cdot [\min(E_i,E_j)]^{2-\alpha}$, $0 <
\alpha \le 2$), but not the $k_t$ algorithm.

So far we have examined the flavour algorithm just for $e^+e^-$
collisions. In hadron-hadron collisions (and some DIS contexts) a
longitudinally invariant algorithm is used, in which the distance
measure $d_{ij}^{kt}$ is obtained by removing the $Q^2$ normalisation,
taking $E_i \to k_{ti}$ and $2(1-\cos\theta_{ij}) \to
\Delta\eta_{ij}^2 + \Delta \phi_{ij}^2$, and by introducing an
additional beam distance measure $d_{iB} = k_{ti}^2$. For the flavour
algorithm, the kinematic modifications are identical, however the beam
distance measure that is introduced is more complex:
\begin{equation}
  \label{eq:diB-flavour}
  d_{iB}^{(F)} = \left\{
    \begin{array}[c]{ll}
      \max(k_{ti}^2, k_{tB}^2(\eta_i))\,, & \quad\mbox{$i$ is flavoured,}\\
      \min(k_{ti}^2, k_{tB}^2(\eta_i))\,, & \quad\mbox{$i$ is flavourless,}
    \end{array}
  \right.
\end{equation}
where one uses a beam hardness as a function of rapidity,
$k_{tB}^2(\eta)$, defined in detail in reference~\cite{BSZFlavour}. It
is more complex to illustrate the IR safety in hadron-hadron
collisions than in $e^+e^-$ because hadron-collider fixed-order (NLO)
programs do not currently provide any information on parton flavour.
So instead we take HERWIG parton shower events, and look
(fig.~\ref{fig:y3dists}b) at the proportion of events in which the
reconstructed jet flavours fail to correspond to the original $2\to2$
event jet flavours before showering, again as a function of the event
softness. All algorithms show a failure rate that vanishes as $y_3
\to0$ because in that limit all parton-showering is forbidden, however
the much faster vanishing for the flavour algorithms is a sign of
their IR safety.  One can also impose \emph{blandness} of
recombinations,\cite{CKKW} \ie forbid multi-flavoured recombinations
(\eg $uu$, $u\bar d$). This has no effect on the $y_3$ dependence of
the failure rate, but improves the overall normalisation.

There are various applications of jet flavour algorithms. A number of
theory calculations (\eg CKKW matching~\cite{CKKW} or hadron-collider
resummations~\cite{BSZhadron}) have a need to project some high
fixed-order calculation onto a flavour channel of lower order so as
to match with an appropriate parton-shower or resummation based on
that lower order.

At hadron level the use of the flavour algorithm is hampered by the
need to know the flavour of every object, making it inappropriate for
experimental discrimination of quark-like versus gluon-like
hadron-level jets. However there does exist one important hadron-level
application: heavy-flavour jets. It is feasible
experimentally~\cite{CDFflav} to identify all heavy-flavour hadrons in
an event. Therefore one can use the flavour algorithm treating only
heavy-flavour objects as flavoured. The infrared safety of the
algorithm means that the heavy-flavoured jet distribution will then be
free of any logarithms of $P_t/m_H$ (other than those in the
heavy-flavour parton distribution), where $P_t$ is the jet transverse
momentum and $m_H$ the heavy-hadron
mass. Furthermore, neglecting power suppressed terms $\sim
m_H^2/P_t^2$, it can be calculated using a normal light-flavour NLO
program~\cite{NLOJET} (extended to provide access to the flavour
information) rather than a dedicated heavy-flavour program.\cite{MNR}

The fact that the new algorithms for safe heavy-flavour jets lead to
far fewer logs of $P_t/m_H$ than currently used heavy-jet
definitions,\cite{FMjets} and that light-flavour NLO programs allow
the remaining logs to be resummed in the PDFs, suggests that
heavy-flavour jet distributions can be predicted significantly better
for the new algorithms, probably reducing the $40-60\%$ uncertainty
typical of current heavy flavour NLO calculations down to the
$10-20\%$ typical of inclusive light-jet distributions at NLO.

\section*{Acknowledgments}
This work is the fruit of collaboration with Andrea Banfi and Giulia
Zanderighi and has been supported in part by grant ANR-05-JCJC-0046-01
from the French Agence Nationale de la Recherche.

\end{document}